\documentclass[twocolumn,showpacs,preprintnumbers,amsmath,amssymb]{revtex4}

\usepackage{graphicx}
\usepackage{amssymb}
\usepackage{hyperref}

\begin{document}

\title{Unstable Shastry-Sutherland phase in Ce$_{2}$Pd$_{2}$Sn}

\author{J.G. Sereni$^{1,*}$, M. Gomez-Berisso$^{1,*}$, A. Braghta$^{2,3}$, G. Schmerber$^3$ and J.P. Kappler$^3$}
\address{$^1$ Div. Bajas Temperaturas, Centro At\'omico Bariloche (CNEA), 8400 S.C. Bariloche, Argentina\\
$^2$ D\'epartement de Physique, Universit\'e de Guelma, 24000 Guelma, Algeria\\
$^3$ IPCMS, UMR 7504 CNRS-ULP, 23 rue de Loess, B.P. 43 Strasbourg
Cedex 2, France}

\date{\today}

\begin{abstract}

{Thermal ($C_P$), magnetic ($M$ and $\chi_{ac}$) and transport
($\rho$) measurements on Ce$_{2}$Pd$_{2}$Sn are reported. High
temperature properties are well described by the presence of two
excited crystal field levels at $(65\pm 5)$K and $(230\pm 20)$K,
with negligible hybridization (Kondo) effects. According to
literature, two transitions were observed at $T_M=4.8$K and $T_C
=2.1$K respectively. The upper transition cannot be considered as a
standard anti-ferromagnetic because of frustration effects in a
triangular network of Ce-atoms and the positive sign of the
paramagnetic temperature $\theta_P^{LT} = 4.4$K. The nature of the
this intermediated phase is described accounting for the formation
of ferromagnetic (F) Ce-dimers disposed in a quasi-2D square
lattice, resembling a Shastry-Sutherland pattern. According to
hysteretic features in $\rho(T)$ and $\chi_{ac}(T)$, the lower
F-transition is of first order, with $C_P(T<T_C)$ revealing a gap of
anisotropy $E_g\approx 7$K.}

\end{abstract}

 \maketitle

%\keywords{Keywords: Cerium Intermetallics, Critical Points,
%Magnetic Phase Diagrams}
%\end{frontmatter}
\section{Introduction}

The competition between different phases is resolved according to
thermodynamic laws, being the configurational entropy an important
factor to define the state of lower free energy ($G$) as the
temperature is reduced. However, complex or meta-stable phases may
occur since roughness in the $G$ function of real systems may
produce relative minima which become relevant. At the transition
temperature, where the $G$ minimum broadens according to the
Ginsburg-Landau theory, the effect of exotic minima is favored.
Within that context, there are non-trivial types of order, like
frustration, incommensurability, glasses or even superconductivity
related to quantum critical points, which may occur when the system
does not accedes directly to an absolute minimum of energy.

The search of experimental examples where these conditions are
realized is an important task in current investigations. The highest
probability to find the mentioned conditions occurs in systems
exhibiting multiple phase transitions because that situation reveals
the competition between different configurations of comparable
energy. According to literature, the Rare Earth based ternaries with
the formula R$_{2}$T$_{2}$X are suitable candidates for such a
purpose, as it was demonstrated by the study of the
Yb$_{2}$Pd$_{2}$(In,Sn) system \cite{Bauer05}. In this In/Sn doped
compound, the stoichiometric limits are non-magnetic heavy Fermions
whilst incipient magnetic order shows up at intermediate In/Sn
substitution.

Crystal chemistry and magnetic properties of R$_{2}$T$_{2}$X
compounds (with R=Ln \cite{Hulliger95,Giovannini98} and Ac
\cite{Peron93}, T=Transition Metals of the VIII group
\cite{Hulliger95,Gordon95,Kacharovsky} and X=early $p$-metals
\cite{Gordon95}) have been investigated over the past two decades
motivated by their peculiar crystalline structure and magnetic
behaviors. Their tetragonal Mo$_{2}$FeB$_{2}$-type structure
\cite{Peron93} is strongly anisotropic and can be described as
successive `T+X' (at z=0) and `R' (at z=1/2) layers.

Ce$_{2\pm x}$Pd$_{2\pm y}$X$_{1\mp z}$ (with $x+y+z=0$) show an
extended range of solubility. This favors that different types of
magnetic configurations with similar energy compete for the
formation of different phases. In fact, by tuning small
excess/deficit of the components ($x, y, z$) one can drive these
compounds between Ferromagnetic (F) and Antiferromagnetic (AF)
order. In Ce$_{2\pm x}$Pd$_{2\pm y}$In$_{1\pm z}$ alloys
\cite{Giovannini98} for example, the existence of two magnetic
branches (c.f. F- and AF-) was determined in the ternary phase
diagram.

In the case of Ce$_{2}$Pd$_{2+x}$Sn$_{1-x}$ stannide
\cite{Fourgeot96} two transitions at $T_{N}=4.7$K and $T_{C}=3.0$K
were reported. Neutron diffusion experiments \cite{Laffarge96}
revealed a modulated character of this intermediate phase, with
the local moments pointing in the `c' direction and an
incommensurate propagation vector $[qx]$ changes from 0.11 (at
4.2K) to 0.077 (at 2.8K). At that temperature it suddenly drops to
$[qx=0]$, becoming ferromagnetic.

Despite of these results there are some contradictory features in
the low temperature behavior of this compound. For example, it is
unlikely to expect a standard AF behavior in a triangular network of
magnetic atoms under geometrical frustration constrains.
Furthermore, the reported value of the paramagnetic temperature
$\theta_P^{LT}$ is positive, and practically coincide with upper
magnetic transition temperature \cite{Braghta}. These properties
impose a revision of the proposed AF character of that phase. For
that reason we will label the upper transition as $T_M$ instead of
$T_N$ like proposed in the literature.

In this paper we present a thorough investigation on thermal,
magnetic and transport properties between and room temperature on
Ce$_{2}$Pd$_{2}$Sn. These results prove the local character of the
Ce moments, allow to gain insight into the knowledge of the nature
of intermediate phase at $T_M>T> T_C$ and the ferromagnetic
character of the ground state.

%%%%%%%%%%%%%%%%%%%%%%%%%%%%%%%%%%%%%%%%%%%%%%%%%%%
\section{Experimental details and results}
Since the magnetic properties of these compounds show a strong
dependence on composition, we have chosen for this study a nearly
stoichiometrc sample with actual composition
Ce$_{2.005}$Pd$_{1.988}$Sn$_{0.997}$ (after SEM/EDAX analysis).
Details of sample preparation were described in a previous paper
\cite{Braghta}. Lattice parameters study confirmed the single
phase character of the sample with Mo$_2$FeB$_2$-type structure.
The respective lattice parameters are: $a=7.765\AA$ and
$c=3.902\AA$, with a $c/a=0.5026$ ratio.

Electrical resistivity was measured between 0.5K and room
temperature using a standard four probe technique with an LR700
bridge. DC-magnetization measurements were carried out using a
SQUID magnetometer operating between 2 and 300K, and as a function
of filed up to 5T. For AC-susceptibility a lock-in amplifier was
used operating at 1.28KHz, with an excitation field of 1Oe on
compensated secondary coils in the range of 0.5 to 10K. Specific
heat was measured using the heat pulse technique in a
semi-adiabatic He-3 calorimeter in the range between 0.5 and 20K,
at zero and applied magnetic fields up to 4T.\\

\begin{figure}
\begin{center}
\includegraphics[angle=0,width= 0.5 \textwidth] {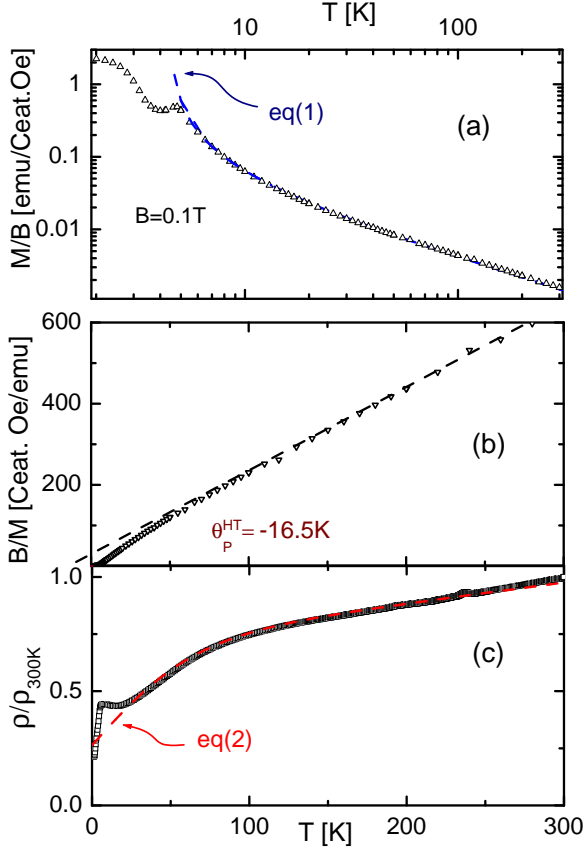}
%{/home/jsereni/papers/publicaz02/pdnial3/textos/F1latpar}
\end{center}
\caption{High temperature properties of Ce$_{2}$Pd$_{2}$Sn. (a)
thermal dependence of field normalized magnetization $M/B$ in a
double logarithmic representation. (b)Inverse magnetization $B/M$,
with B=0.1T. Straight line represents the extrapolation from high
temperature to extract $\theta_P^{HT}$. (c) Temperature dependence
of normalized electrical resistivity $\rho(T)/\rho_{300K}$. Dashed
curves in (a) and (c) are fits using eq.(1) and (2) respectively,
see the text.} \label{F1RhoMag}
\end{figure}

%%%%%%%%%%%%%%%%%%%%%%%%%%%%%%%%%%%%%%%%%%%%%%%%%%%%%%%%%%%
\subsection{High temperature results} High temperature
DC-magnetization $M/B$ and electrical resistivity $\rho$ results are
presented in Fig.1. The former is depicted as $M/B vs. T$ in a
double logarithmic scale (Fig.1a), whilst the latter was normalized
to the room temperature value $\rho_{300K}$ (Fig.1c). Both curves
are in good agreement with results obtained on similar samples
\cite{Fourgeot96,Gordon96}. Fig.1b shows the inverse of the
magnetization $B/M$ used to extract the high temperature effective
moment $\mu_{eff}^{HT}$ and paramagnetic Curie-Weiss temperature
$\theta_P^{HT}$.

$M(T)$ was measured up to room temperature with an applied field
$B=0.1$T. At $T_M=4.8$K, it shows a cusp which was previously
taken as an indication for an AF character of that transition.
However, the divergent increase of $M/B(T\to T_I)$ with
$T_I=4.4$K, makes that argument uncertain. The analysis of these
results can be done applying a simple formula \cite{PdRh}:
\begin{equation}
M/B =\sum_0^2 \mu_i^2*exp(-\Delta_i/T)/ [(T-T_I)*Z]
\end{equation}
where $\mu_i$ are the effective moments of the ground state (i=0)
and excited CF levels (i=1,2 respectively). Such a good fit in
that wide range of temperature, with an equation only taking into
account the Boltzmann thermal occupatio of the excited CF levels
indicates that those levels very well defined in energy, excluding
any significant "4f-band" hybridization effect.

The values obtained by applying this formula are: $\mu_0=1.7\mu_B$
and $\mu_1=1.9\mu_B$ and $\mu_2=2.48\mu_B$, and the respective CF
splitting: $\Delta_1=65\pm 4$K and $\Delta_2=230\pm 20$K. Since at
room temperature the upper CF level is not jet fully occupied, the
value for a $J=5/2$ moment is not reached.

The inverse of the susceptibility $B/M$ shows a Curie-Weiss
behavior at high temperature (i.e. $T \geq 80$K, see Fig.1b) with
an extrapolation for the paramagnetic Curie-Weiss temperature:
$\theta_P^{HT}=-16.5$K, in agreement with the literature
\cite{Fourgeot96}. However, below $T\approx 80$K a negative
curvature takes over as the thermal population of the excited CF
levels decreases. This results in a positive value of the low
temperature extrapolation: $\theta_P^{LT} \to 4.4$K (see Fig.2a),
in coincidence with the divergency at $T\to T_I$ in eq(1). The
proximity of $\theta_P^{LT}$ to $T_M$ is an evidence of a
competition between different types of interactions, since the
former is related to a F- divergence and the later to an AF-like
cusp in $M(T)$.

\begin{figure}
\begin{center}
\includegraphics[angle=0,width= 0.5 \textwidth] {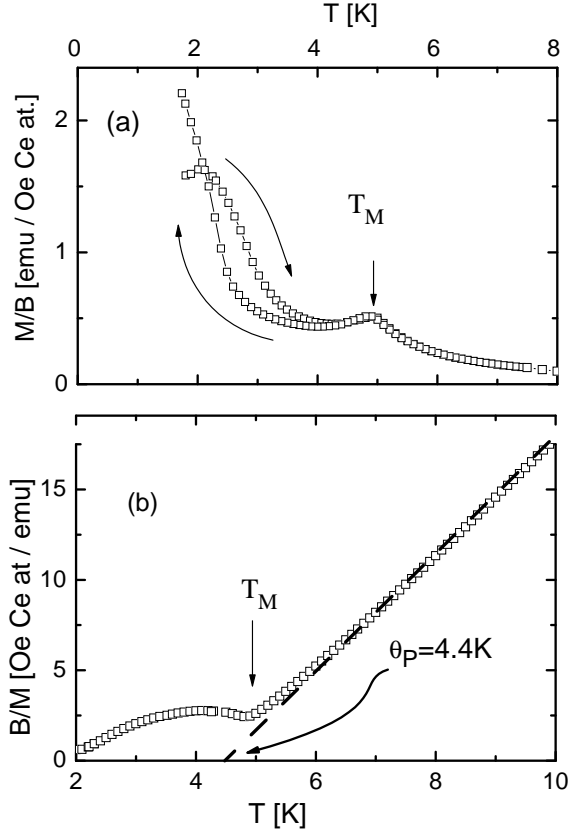}
%{/home/jsereni/papers/publicaz02/pdnial3/textos/F1latpar}
\end{center}
\caption{(a) Comparison of Magnetization results in field (B=5mT)
cooling (FC) and heating (after ZFC) processes. (b) Low temperature
inverse magnetization $B/M$, with B=0.1T. Dashed line represents the
extrapolation from high temperature to obtain $\theta_P^{LT}$.}
\label{F2RhoField}
\end{figure}

$\rho(T)$ dependence can be described following the Matthiessen
criterion: $\rho(T)=\rho_0+\rho_m(T)+\rho_{ph}(T)$, where $\rho_0$
is the residual resisitivity, $\rho_m(T)$ the magnetic
contribution and $\rho_{ph}(T)$ the phonon contribution to
electronic scattering. Due to the mentioned negligible
hybridization effects on the excited CF levels $\rho(T)$ is very
well described by the simple expression \cite{Pd2Al3}, as shown in
Fig.1c:
\begin{equation} \rho(T)= a + b * tanh(T/D) + c * T
\end{equation}
where $a=\rho_0$, $b * tanh(T/D) = \rho_m$ and $c * T =
\rho_{ph}(T)$. The parameter $D$ indicates the characteristic energy
of the dominant scattering center, which in this case corresponds to
the first excited crystal field (CF) level $\Delta_1=63$K. The
second CF excited level can be estimated at $\Delta_2 > 200$K. These
values are in good agreement with those extracted from $M/B(T)$.\\

\subsection{Low temperature results ($T<20$\,K)}
 As indicated by $M(T)$ results, the
low temperature behavior of this compound is rather complex. Hence,
some previous considerations are required to better understand the
experimental results presented in this section. Since the AF
molecular field (with $\theta_P^{HT}=-16.5$K) turns to ferromagnetic
($\theta_P^{LT} \to 4.4$K) below about 20K, one may distinguish a
correlated-paramagnetic region between $20K>T>T_M$ from the
canonical paramagnetism above that temperature. As mentioned before,
despite of a cusp at $M(T_M)$ the upper transition is strongly
affected by those F- type correlations. This situation persists in
the intermediate phase ($T_M>T>T_C$)and it is only below $T_C$ that
the stable F- ground state is established.

\begin{figure}
\begin{center}
\includegraphics[angle=0,width= 0.5 \textwidth] {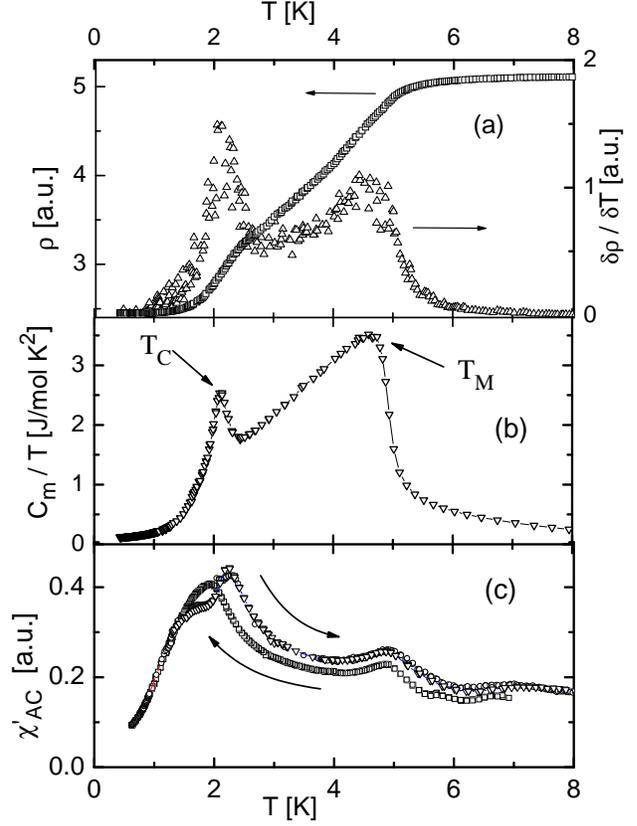}
%{/home/jsereni/papers/publicaz02/pdnial3/textos/F1latpar}
\end{center}
\caption{(a) Low temperature electrical resistivity (left axis) and
its temperature derivative (right axis). (b) Magnetic contribution
to specific heat divided by temperature showing the upper
(modulated) $T_M$ and the lower (F) transitions. (c)
AC-susceptibility as a function of temperature in cooling and (two
runs) on heating procedures.} \label{F3RhoCmChi}
\end{figure}

Between $20K>T>T_M$, $\rho(T)$ shows a slight increase which could
be attributed to Kondo effect (see Fig.1c). Such an increase of
$\rho(T\to T_M)$ might be attributed to Kondo effect, however, it is
also observed above Ferromagnetic transitions, like in GdPt
\cite{GSal86}. In that case, the increase of $\rho(T\to T_C)$ arises
from F-fluctuations, precursors of the magnetic transition. Below
$T_M$, $\rho(T)$ felts down without and sign of AF gap opening at
the transition. At $T=T_C$, hysteretic effects are observed, which
are practically suppressed by a field of nearly $0.2T$ (see Fig.2b).

Specific heat, $C_P(T)$, also shows an increasing contribution
between $20>T>T_M$ (see the Fig.3c) originated in magnetic
correlations precursors of $T_M$ transition. Since this $C_P(T>T_M)$
tail contains important information about the nature of the related
transition, it will be analyzed in detail in the discussion
section.\\

%%%%%%%%%%%%%%%%%%%%%%%%%%%%%%%%%%%%%%%%%%%%%%%%%%%%%%%%%%%%%%%%%%%%%%%
{\bf Modulated phase ($T_M>T>T_C$)}. $M(T)$ is certainly the most
sensitive parameter to F-correlations. The low field (B=5mT)
magnetization results shown in Fig.3a confirm the ambiguous behavior
of the intermediate phase. The difference between cooling
($T\downarrow$) and heating ($T\uparrow$) processes are evident,
especially below $T_I=4.4K (=\theta_P^{LT})$ which can be considered
as a temperature of irreversibility. This means that the AF
character proposed in the literature owing the modulated magnetic
structure is never properly established.

Both magnetic transitions are determined from $C_m/T(T)$
measurements by the respective jumps at $T_M=4.8$K and $T_C=2.1$K
respectively, as shown in Fig.3b. $C_m/T(T)$ is obtained after
phonon subtraction extracted from La$_{2}$Pd$_{2}$Sn as:
$C_m/T=C_P/T-C_ P/T(La_{2}Pd_{2}Sn)$. The $C_m(T)$ dependence
between $T_M \geq T \geq T_C$ does not fit into the expected for a
canonical AF-, c.f. $C_m(T)\propto T^3$. Instead, the observed
dependence: $C_m(T)= 0.5 T^{2.25}$ is closer to that of a strongly
anisotropic ferromagnet with $C_m(T)\propto T^{2.5}$
\cite{spinwaves}.\\

Ac-susceptibility $\chi_{ac}$ measurements, presented in Fig.3c,
confirm these findings with an increasing signal below 4K and a
maximum at $T_C$ which depends on cooling or heating process and
confirm hysteretic behavior.

%%%%%%%%%%%%%%%%%%%%%%%%%%%%%%%%%%%%%%%%%%%%%%%%%%%
{\bf Ferromagnetic phase ($T>T_C$)}. The hysteretic behavior between
$T\downarrow$ (with $\chi_{ac}^{max}$ at $T\approx 1.9$K) and
$T\uparrow$ (with $\chi_{ac}^{max}$ at $T\approx 2.2$K) indicates
the first order character of the ferromagnetic transition. The lower
limit of this hysteretic region is observed $\approx 1$K. In the
heating process, a shoulder is observed in $\chi_{ac}(T\uparrow)$ at
$T\approx 1.5K$. This anomaly can be attributed to a very small
amount of ferromagnetic Ce$_3$Pd$_5$ \cite{Kappler}, which is not
even detected in $C_P(T)$ nor in $\rho(T)$ measurements.

\begin{figure}
\begin{center}
\includegraphics[angle=0,width= 0.5 \textwidth] {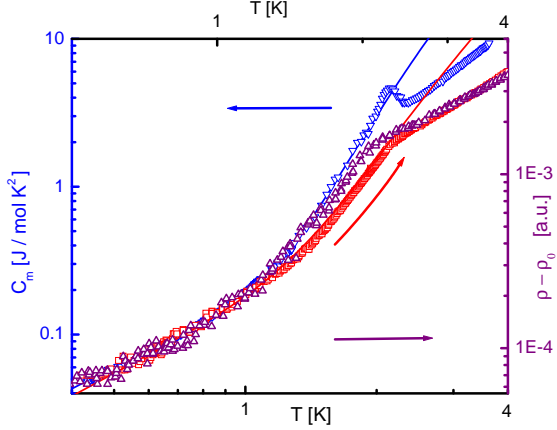}
%{/home/jsereni/papers/publicaz02/pdnial3/textos/F1latpar}
\end{center}
\caption{Comparison of $T$ dependencies of magnetic specific heat
(left axis) and electrical resistivity (right axis) of the F-phase.
Cooling and heating runs of resistivity show a hysteresis at $T_C$.}
\label{F4CmRho}
\end{figure}

Also the temperature dependence of the electrical resistivity
shows hysteretic behavior between cooling and heating around $T_C$
as it can be seen in the inset of Fig.2. From specific heat
results, the ferromagnetic transition occurs at $T_C=2.1$K with
sharp anomaly characteristic of first order type, which involve a
small amount of latent heat.

Within the F- phase, both $C_m(T)$ and $\rho_m(T)$ are very well
described by their respective F- temperature dependencies
containing an exponential factor due to the presence of a gap in
the magnon spectrum. Those functions are \cite{Sundstrom}:
\begin{equation}
C_m(T)=\gamma T + T^{3/2} (a+b \exp(-E_g/T))
\end{equation}
with $\gamma=7 mJ/molK^2$ confirming the absence of Kondo effect,
and $E_g=7.5$K. The electrical resistivity is well described by
\cite{Madeiros}:
\begin{equation}
\rho(T)\propto T^{1/2}\exp(-E_g/T)(1+cT/E_g+d(T/E_g)^2)
\end{equation}

with a similar value for the gap: $E_g=7.2$K. The presence of this
gap is a clear indication of the strong anisotropy of this system.

\begin{figure}
\begin{center}
\includegraphics[angle=0,width= 0.5 \textwidth] {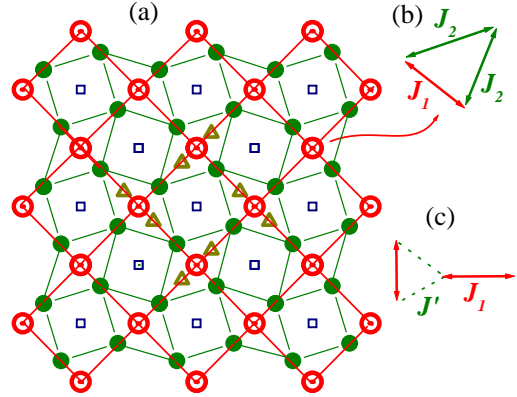}
%{/home/jsereni/papers/publicaz02/pdnial3/textos/F1latpar}
\end{center}
\caption{(a) Mo$_2$FeB$_2$ type structure showing Ce (green
$\bullet$), Sn (blue $\Box$) and Pd (yellow $\bigtriangleup$) atoms.
Superposed is a schematic representation of a square lattice (red
lines) built up by F dimers ($\odot$) with moments pointing in the
'c' direction. (b) Magnetic interactions: $J_1$ (red) and $J_2$
(green), between Ce atoms on the 'ab' plane. (c) Magnetic
interactions $J'$ (green dots) between neighboring dimers.}
\label{F5dimeros}
\end{figure}

%%%%%%%%%%%%%%%%%%%%%%%%%%%%%%%%%%%
\section{Discussion}

The strong local character of the "$4f$" orbital, as a well defined
trivalent Ce atom, is concomitant with the irrelevant hybribization
(or Kondo) effect on ground (GS) and CF excited states. Within such
a scenario, a F- ground state is usually expected for Ce compounds
\cite{Physica215}. Knowing that the twin compound Ce$_{2\pm
x}$Pd$_{2\pm y}$In$_{1\pm z}$ modifies its magnetic structure with a
slight variation of composition \cite{Giovannini98}, one should ask
whether the intermediate phase is an exotic phase appearing under
this special situation of competing exchange interactions. Thus, the
formation of such a phase previous to the stabilization of the F-GS
rises the question about the role of the atomic configuration in
this peculiar crystalline structure together with the consequent
steric distribution of magnetic interactions. Accordingly,
modulated/inconmensurated nature of the propagation vector hints for
an exotic character of that phase. Therefore, its origin should be
searched mainly taking into account its structural
properties rather than only the electronic ones.\\

\begin{figure}
\begin{center}
\includegraphics[angle=0,width= 0.4 \textwidth] {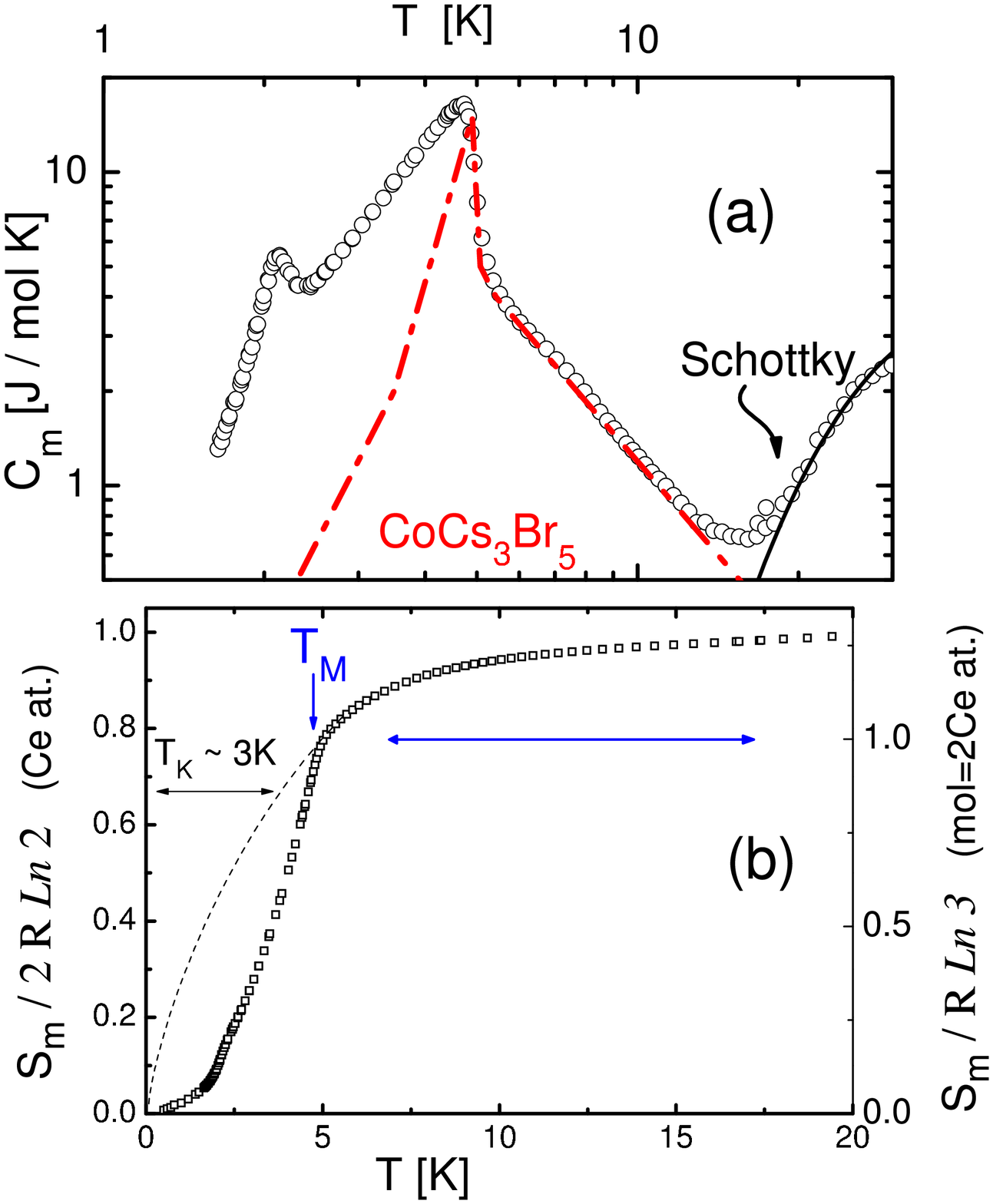}
%{/home/jsereni/papers/publicaz02/pdnial3/textos/F1latpar}
\end{center}
\caption{a) Comparison of measured "molar" specific heat tail at
$T>T_M$ with that of an exemplary system of 2-D Ising S$_{eff}$=1
square lattice CoCs$_3$Br$_5$ \cite{Wielinga}. b) Comparison of
thermal entropy gain between $R Ln 2$ (right axis) and $R Ln 3$
(left axis) scales} \label{F6CompCp&Entropy}
\end{figure}

{\bf Structural Properties}
 The structural disposition of R atoms in these compounds
can be depicted as trigonal and tetragonal prisms, centered around
`T' and `X' atoms respectively, dispose the `R' atoms in non
centro-symmetric positions along the `c' direction. As a consequence
of the trigonal disposition, there a "one to one" correlation of
magnetic properties between Ce$_{2}$T$_{2}$X compounds and the
correspondent CeT binary compounds \cite{Gsch} with CrB- or FeB-type
structures (both structured in trigonal prisms \cite{kiess50}). This
results in an assembled mosaic of rhombuses and squares projected on
the basal plane. While the rhombuses are rotated $\pi$/2 respect to
their four parent nearest-neighbors ($nn$), the squares are rotated
$\pi$/4 among them (see Fig.5), mimetizing a sort of "pinwheels"
centered on the Sn-Sn column.

An important characteristic of this family of compounds is that the
shortest $d_{R-R}$ spacing (i.e. the nearest R-R neighbors, $Rnn$)
can be either in the `c' direction (2 $nn$) or in the basal `ab'
plane (1 $nn$) depending on the `c/a' ratio of the lattice
parameters. Values of $c/a<0.5022$ favors the formation of a `R'
chains parallel to `c', like in $U_{2}Ni_{2}In$ \cite{Peron93}. On
the contrary, for larger 'c/a' values there is only one R-R first
$Rnn$ lying on the basal plane. Their distance is the shorter
diagonal of the rhombus polygon (dashed line in Fig.7) and separates
two triangular prisms disposed in mirror position, centered on the
Pd-Pd column.

In Ce$_2$Pd$_2$Sn, the ratio c/a=0.5026 indicates that the $nn$ is
only one atom lying on the z=1/2 plane. However, the difference
respect to the two next $nn$ (in the "c" direction) is less than
$1\%$: $d_{Ce-1Ce}= 3.882\AA$ and $d_{Ce-2Ce}= 3.902\AA$
respectively. These two Ce-Ce spacings lie within the
$3.7<d_{Ce-Ce}<4.0\AA$ range where F-Ce binaries are placed
\cite{Gsch}, including CePd.\\

{\bf The nature of the modulated phase}. To evaluate the
consequences on these structural characteristics on the magnetic
properties, one have to take into account that the effective RKKY
interaction depends on three parameters:
\begin{equation}
J_{eff}(r)=J_{ex}*\mathbb{S}(\mathbb{S}+1)*f_{RKKY}(r)
\end{equation}
where $J_{ex}$ is the intensity of the exchange interaction between
$R-nn$ atoms, $\mathbb{S}$ represents the magnetic spin and
$f_{RKKY}(r)$ is the RKKY oscillatory function. One have to consider
different values of $J_{ex}$ corresponding to each type of neighbor:
$J_{1}$ and $J_{2}$ for the respective $nn$ (one) and the next $nnn$
(four) neighbors on the `ab' plane, see Fig.5b. According to neutron
measurements, $J_{c}$ connecting Ce atoms with those of the upper
and lower plane is not relevant in this range of temperature.\\

Although this compound shows its upper magnetic transition at
$T_M=4.8K$, the presence of magnetic correlations is observed at
much higher temperatures in the tail of $C_m(T>T_M)$ and the
increase of $\rho(T<15K)$. These signs of incipient magnetic
correlations within the paramagnetic phase contains important
information about the nature of the transition itself since they
contain the contribution of related precursors. Furthermore,
strongly anisotropic or low dimensional systems display a
significant amount of entropy above the transition. The existence of
such a correlated paramagnetic region in the compound under study is
confirmed by the fact that it has to be heated at least up to
$\approx 20$K for a ZFC process, otherwise traces of remand magnetic
contributions are detected afterwards.

A deeper analysis of $C_m(T)$ shows that the temperature dependence
at $T>T_M$ is $C_m\propto 1/T^2$, in coincidence with the specific
heat of CoCs$_3$Br$_5$ \cite{Wielinga} (see Fig.6a), which is an
archetype for a quadratic spin $\mathbb{S}_{eff}=1$ square lattice.
Notice that not only the temperature dependence coincides but also
their respective absolute values, provided that for Ce$_2$Pd$_2$Sn
$C_m$ is taken in "mol" units (i.e. 2 Ce atoms). Notice that the
transition temperature of the model compound is scaled to compound
under study. The jump of the specific heat also points in that
direction, since in mean field theory the jump of the specific heat
at a magnetic transition is given by: $\Delta
C_m=2.5R[(2\mathbb{S}_{eff}+1)^2-1]/[(2\mathbb{S}_{eff}+1)^2+1]$
\cite{SpHt}, which for $\mathbb{S}_{eff}=1/2$ is $\Delta C_m =
1.5$R, whilst for $\mathbb{S}_{eff}=1$ it becomes $\Delta C_m = 2$R
like the measured value.

The magnetic entropy, computed as $\Delta S_m=\int C_m/T dT$,
reaches the expected value $\Delta S_m = 2RLn2$ for a doublet ground
state at around 15K (the pre-factor 2 corresponds to two magnetic
atoms per formula unit). As depicted in Fig.5b, the entropy gain
reaches $80 \%$ of that value at $T=T_M=4.8$K. This is a direct
consequence of the low value of $T_K\approx 3.0$K, which can be
evaluated from $\Delta S_m(T)$ by applying the Desgranges-Schotte
\cite{Desgr} criterion of $\Delta S_m(T=T_K) \cong 2/3 RLn2$ for a
non-ordered system.

The facts that, in the comparison with CoCs$_3$Br$_5$, the $C_m(T)$
of Ce$_2$Pd$_2$Sn has to be computed as "per mol" (i.e. = 2Ce at.)
and that the reference compound has an effective spin
$\mathbb{S}_{eff}=1$, provide the keys to understand the physics of
our compound. They suggest the formation of magnetic dimers built up
through the $J_1$ interaction. Since the $\mathbb{S}_{eff}=1$
momentum has three quantum projections, the measured entropy has to
increase accordingly. In fact, the 80\% of $\Delta S_m = 2RLn2$ (per
Ce at.) collected up to $T=T_M$ corresponds to $0.5 RLn3$ (i.e 'per
mol'), as depicted on the right axis of Fig.5b. The 20\% of the
remanent value is related to the entropy relaxed in the $C_m(T>T_M)$
tail which arises from the difference between: $(Ln2-0.5Ln3)/Ln2)$.

The consequent structural loci of those dimers in a square lattice
is sketched in Fig.4a, where the $\mathbb{S}_{eff}=1$ moments are
depicted in the center of the $d_{Ce-1Ce}$ spacing as $\odot$. This
structural configuration of dimers is topologically equivalent to
the Shastry-Sutherland systems \cite{Shastry} proposed for AF dimers
and inter-dimer interactions, i.e. AF $J_1$ and $J'$. As it was
pointed out by theoretical calculations \cite{Miyahara} Such
combination of interaction can lead to a stable AF-GS, recently
realized in Yb$_2$Pt$_2$Pb single crystals \cite{Kim}, which
undergoes a slight shift of Pt atoms that results in two kinds of
Yb-Pt tetrahedra.

Comparing these Yb and Ce isotypic compounds, we can remark the
heavy fermion character of Yb$_2$Pt$_2$Pb, with $\gamma =
311$\,mJ/mol K$^2$ \cite{Kim}, whereas the Ce one shows a record low
value of $gamma = 7$\,mJ/mol K$^2$. The hybridization effect may
explain the lower entropy value of Yb$_2$Pt$_2$Pb at $T=T_N$
[$\Delta S_m(T_N) = 0.58 RLn 2$] compared with that of
Ce$_2$Pd$_2$Sn [$\Delta S_m(T_M) = 0.80 RLn 2$]. Among the
similarities, one can mention the similar CF splitting of the first
excited level ($\Delta_1=70$\,K and 65\,K, respectively), though
their $1/chi(T)$ curves show opposite signs. In both compounds, the
extended tails above their respective upper transitions become
significant at about three times $T_N$ ($T_M$). At lower
temperatures, they show an anomaly at nearly one half of the upper
transition. Noteworthy is the fact that, while in Yb$_2$Pt$_2$Pb it
looks like a shoulder (see Fig.5b in \cite{Kim}), it is a first
order transition in Ce$_2$Pd$_2$Sn.

Since in the compound under study $J_1$ is ferromagnetic, and that
situation was not proved theoretically to become a stable GS, our
magnetic scenario seems to be non trivial or perhaps even
meta-stable. Furthermore, taking into account the modulated
character of the intermediate phase, an AF interaction between
dimers (i.e. $J'$ in Fig.5c) can be expected. Particularly, in the
case of Ce$_2$Pd$_2$Sn, the low temperature magnetic properties
suggest that the proposed Ce-Ce magnetic dimerization arises once
$\Delta S_m$ becomes $< R Ln2$ (at $T\leq 20$K as suggested in
Fig.6a) when the paramagnetic degrees of freedom of the doublet GS
start to condense. Coincidentally, the straight line extrapolating
$1/\chi$ to $\theta_P^{LT} = 4.4$\,K is well defined up to that
temperature. However, when those dimers start to interact AF to each
other via $J^{'}$ (as expected from the cusp of $M(T)$ at $T_M$),
the long range order parameter cannot stabilize in a simple manner.
Notice that only the couplings between $nn$ dimers is depicted in
Fig.5c for simplicity, whilst the Hamiltonian proposed in
ref.\cite{Miyahara} involves a set of next-$nn$ couplings. This
apparent intrinsic instability of the intermediate phase (probably
originated in frustration effects) ends at the lower transition
$T_C=2$K, where a 3D F-GS takes over.

\section{conclusions}

These experimental results confirm Ce$_2$Pd$_2$Sn as one of the
scars examples of ferromagnetic ground state among Ce intermetallic,
with very stable magnetic moments and practically no traces of Kondo
effect in the ground and excited CF states.

At intermediate temperatures ($T<20$K) magnetic interactions develop
with an excitation spectrum mimetizing a quasi-2D square lattice
$\mathbb{S}_{eff}=1$ model. This property, the entropy value at
$T_M$ [$\Delta S_m(T_M) = RLn 3$] and the structural morphology
suggest the progressive formation of Ce-Ce ferromagnetic dimers
within a pattern resembling that proposed by Shastry-Sutherland. The
F character of $J_1$ is recognized from the positive value of
$\theta_P^{LT}$ and the entropy related to a three fold degenerate
state, whereas the AF character of the exchange $J^{'}$ between
dimers explains the cusp of $M(T)$ at $T_M=4.8$\,K. Nevertheless,
this interaction is not able to stabilize the intermediate phase
($T_M \geq T \geq T_C$) leading the system to another phase
transition.

The comparison with isotypic Yb$_2$Pt$_2$Pb and model predictions
suggest that no stable GS is reached in a combination of F-dimers
($J_1$) and AF exchange ($J^{'}$) among them, at least in this
intermetallic compound. Nevertheless, an exotic phase with a non
trivial order parameter compete in energy within a short range of
temperature till the 3D F-GS takes over. At the Curie temperature
$T_C =2.1$K a first order transition is observed in $C_P(T)$,
$\rho(T)$ and $\chi_{ac}$, including hysteretic features. Below
$T_C$, the $C_P(T)$ dependence reveals a gap of anisotropy
$E_g\approx 7$K in the spectrum of mangnons.

Further measurements are in progress to investigate the magnetic
phase diagram of Ce$_2$Pd$_2$Sn and the effect of a slight deviation
from stoichiometry.

\section*{Acknowledgments}
We are grateful to Dr.C.Geibel and M. Giovannini for clarifying
discussions, and to Ms. Ogando for her participation in magnetic
measurements. This work was partially supported by PIP-6016
(CONICET) and Secyt-UNC project 6/C256.\\
* J.G.S. and M.G-.B. are members of the Consejo Nacional de
Investigaciones Cient\'ificas y T\'ecnicas and Instituto Balseiro
(UN Cuyo) of Argentina.

%\subsection*{Corresponding author: jsereni@cab.cnea.gov.ar (J.G.
%Sereni)}


\begin{thebibliography}{99}

\bibitem{Bauer05} E. Bauer et al., J. Phys.: Condens Matter 17
(2005) S999.

\bibitem{Hulliger95} F. Hulliger, J. Alloys \& Comp. 221 (1995) L11.

\bibitem{Giovannini98} M. Giovannini et al., Phys. Rev. B 61 (2000) 4044.

\bibitem{Peron93} M.N. Peron et al., J. Alloys and Comp. 201 (1993) 203.

\bibitem{Gordon95} R.A. Gordon et al., J. Alloys and Comp. 224 (1995) 101.

\bibitem{Kacharovsky} D. Kacharovsky et al., Phys. Rev. B 54 (1996) 9891.

\bibitem{Fourgeot96} F. Fourgeot et al., J. Alloys and Comp. 238 (1996) 102.

\bibitem{Laffarge96} D. Laffarge et al., Solid State Commun. 100
(1996) 575.

\bibitem{Braghta} A. Braghta et al., J. Magn. and Magn. Materials
320 (2008)1141.

\bibitem{Gordon96} R.A. Gordon and F.J. Di Salvo, J. Alloys and Comp. 238 (1996) 57.

\bibitem{Pd2Al3} J.G. Sereni et al., J. Phys.: Condens. Matter 18 (2006) 3789.

\bibitem{PdRh} J.G. Sereni et al., Phys. Rev. B 75 (2007) 024432.

\bibitem{GSal86} J.C. Gomez-Sal et al., Solid State Commun. 59
(1986) 771.

\bibitem{spinwaves} A.I. Akhiezer et al., in {\it Spin waves},
North-Holland Series in Low Temperature Physics, Ed. C.J. Gorter,
Vol I, ch. 6, p. 201, 1968, N-H Pub. Co.

\bibitem{Kappler} J.P. Kappler, M.J. Besnus, P. Lehmann, A Meyer and J.G. Sereni, J. Less-Comm. Metals 111 (1985) 261.

\bibitem{Sundstrom} L.J. Suntr\"om, in {\it Handbook for Physics and Chemistry of Rare
Earths}, edited by K.A. Gschneidner Jr. and L. Eyring, Chap. 5,
Vol 1 ( Morth-Holland Pub. Co., 1978).

\bibitem{Madeiros} See e.g. S.M Madeiros et al., Physica B 281 (2000) 340 and
M.A. Continentino et al., Phys. Rev. B 64 (2001) 012404.

\bibitem{Physica215} J.G. Sereni, Physica B 215 (1995) 273.

\bibitem{Gsch} see for example: J.G. Sereni, in {\it Handbook for Physics and Chemistry of Rare
Earths}, edited by K.A. Gschneidner Jr. and L. Eyring, Chap. 98, Vol
15 (Elsevier Science B.V, 1991).

\bibitem{kiess50} R. Kiessling, Acta Chem. Scandinavica 4 (1950)
209.

\bibitem{Wielinga} R.F. Wielinga, H.W. Bl\"ote, J.A. Roest, W.J.
Huiskamp, Physica 34 (1967) 223.

\bibitem{SpHt} See e.g. J.G. Sereni, in {\it Encyclopaedia of
Materials, Science Technology}, Ed. K.H. Buschow, Sub.Ed. E.
Gratz, Vol. 5, I-Mag, Elsevier Sci.Ltd. (2001), p. 4986.

\bibitem{Desgr} H.-U. Desgranges and K.D. Schotte, Physics Letters
91A (1982) 240.

\bibitem{Shastry} B.S. Shastry and B. Sutherland; Physica 108B (1981)
1069.

\bibitem{Miyahara} S. Miyahara and K. Ueda; Phys. Rev. Lett. 82.
(1999) 3701.

\bibitem{Kim} M. S. Kim, M.C. Bennett and M.C. Aronson; Phys. Rev.
B 77 (2008) 144425.

\end{thebibliography}
\end{document}